\documentclass[twocolumn]{autart}
\usepackage{natbib}
\usepackage{epsfig}
\usepackage{amsmath}
\usepackage{amssymb}
\usepackage{graphicx}
\usepackage{subfigure}
\usepackage[dvipsnames,usenames]{color}
\usepackage{balance}
\usepackage{multicol}

\begin{document}

\begin{frontmatter}
\title{Distributed tracking control of leader-follower multi-agent systems under noisy measurement \thanksref{footnoteinfo}}

\thanks[footnoteinfo]{This paper was not presented at any IFAC meeting.}
\author[hu]{Jiangping Hu\corauthref{cor}}\corauth[cor]{Corresponding author:  Jiangping Hu. Tel. +86-28-61831590.
Fax +86-28-61831113.}\ead{hjp$\_$lzu@163.com} \; and \;
\author[feng]{Gang Feng}\ead{megfeng@cityu.edu.hk}

\address[hu]{School of Automation Engineering, \\
                     University of Electronic Science and Technology of China, Chengdu 610054, China}
\address[feng]{Department of Manufacturing Engineering and Engineering Management, \\
              City University of Hong Kong, Kowloon, Hong Kong}

\date{}

\begin{keyword}
Multi-agent systems,  leader-follower, velocity decomposition, state
estimation, stochastic noises.
\end{keyword}

\begin{abstract}
In this paper, a distributed tracking control scheme with
distributed estimators has been developed for a leader-follower
multi-agent system with measurement noises and directed
interconnection topology. It is supposed that each follower can only
measure relative positions of its neighbors in a noisy environment,
including the relative position of the second-order active leader. A
neighbor-based tracking protocol together with distributed
estimators is designed based on a novel velocity decomposition
technique. It is shown that the closed loop tracking control system
is stochastically stable in mean square and the estimation errors
converge to zero in mean square as well. A simulation example is
finally given to illustrate the performance of the proposed control
scheme.
\end{abstract}

\end{frontmatter}

\section{Introduction}
In recent years  cooperative distributed control of multi-agent
systems has been a research focus in control community. An important
control strategy among many others is the leader-following
coordination among a team of agents. The leader-follower approach
has been widely used in many practical applications such as
formation control in robotic systems (\cite{wang, das}),  unmanned
aerial vehicle (UAV) formation (\cite{vane,anders}), target tracking
in sensor network (\cite{gupta,hu}), and so on.

The major issues addressed in the study of leader-follower
multi-agent systems include the varieties of topological consensus
conditions (\cite{jad,ren}), the roles of multiple leaders in
guiding the followers  (\cite{lin,shi}), the time-delayed control
design (\cite{hu07,lin08}), and the distributed estimation
strategies (\cite{fax,hong06}). A common feature of these existing
works on distributed control for leader-follower multi-agent systems
is that the measurement noises are not considered. However in
practice,the measurements and information communication are always
subject to noises and/or perturbations, such as sensor noise,
channel fading, quantization errors, etc. More recently consensus
control problems with measurement noises have been studied in
\cite{li09} and \cite{huang} with the fixed and undirected network
topology.  \cite{huang} proposed a consensus control for a
leaderless multi-agent system with the first order discrete-time
dynamics under noisy measurements and proved that the average
consensus can be achieved in the sense of mean square by introducing
a decreasing gain if the network topology is a strongly connected
circulant graph. \cite{li09} extended the result to the first order
continuous-time average consensus problem and obtained a sufficient
and necessary condition. To the best of our knowledge, there is no
report in open literature on design of distributed control for a
leader-follower multi-agent system with measurement noises and
time-varying directed interconnection topology.

In this paper, we will consider a distributed control design for a
leader-follower multi-agent system under partial and noisy
measurements and time-varying directed network topology. A novel
velocity decomposition technique, inspired by the stochastic
approximation approach (\cite{neve}), and a distributed estimation
algorithm for the velocity of the active leader have been proposed
to deal with those partial and noisy measurements. It has been shown
that the estimation is convergent in mean square and the resulting
closed loop control system is stochastically stable in mean square.

The remainder of this paper is organized as follows. In Section
\ref{form}, some concepts in algebraic graph theory are briefly
reviewed and a leader-following problem is formulated. In Section
\ref{design}, a tracking control along with a  distributed
estimation algorithm is  firstly designed for the leader-follower
multi-agent system based on the velocity decomposition technique.
Then the stochastic stability of the closed-loop tracking error
system is analyzed under switched directed topology. A numerical
example is given to illustrate the distributed tracking control for
the leader-follower multi-agent system in Section \ref{simu}.
Finally, some concluding remarks and future research directions are
given in Section \ref{con}.

Throughout this paper, we will use the following notations. $I$
denotes an appropriate dimensioned identity matrix; $\mathbf{1}$
denotes a column vector with all ones.  For a given  matrix $A$,
$A^T$ denotes its transpose; $tr(A)$ its trace; $\|A\|$ its
Frobenius norm; $\lambda_{\rm max}(A)$ and $\lambda_{\rm min}(A)$
its maximum and minimum eigenvalues respectively. For a given set
$S$, $\chi_S$ denotes the indicator function of $S$. $E[\cdot]$ is
the expectation operator; $col(\cdot)$ denotes the concatenation.
For any given real numbers $a$ and $b$, $a\wedge b$ denotes
$min\{a,b\}$.

\section{Problem formulation}\label{form}

\subsection{Preliminaries}

In order to describe the interconnection topology of a
leader-follower multi-agent system, we need to introduce some
preliminaries from algebraic graph theory \cite{cgod}.

Let $\mathcal{G}=(\mathcal{V},\mathcal{E})$ be a \textbf{directed
graph} (or digraph for simplicity) consisting of a finite set of
vertices $\mathcal{V}=\{0,1,...,n\}$ and a finite set of arcs
$\mathcal{E}\subseteq \mathcal{V} \times \mathcal{V}$. The order of
$\mathcal{G}$ is the number of vertices in $\mathcal{G}$ and denoted
by $|\mathcal{G}|$. An arc of $\mathcal{G}$ is denoted by $(i,j)$,
which starts from $i$ and ends on $j$ and represents the information
flow from agent $j$ to agent $i$. A path in $\mathcal{G}$ is a
sequence $i_0, i_1,\cdots,i_q$ of distinct vertices such that
$(i_{j-1},i_j)$ is an arc for $j=1,\cdots, q$. If there exists a
path from vertex $i$ to vertex $j$, we say that vertex $j$ is
reachable from vertex $i$. Furthermore, if there exists a path from
every vertex to vertex $j$, then vertex $j$ is \textbf{a globally
reachable vertex} of $\mathcal{G}$. A digraph $\mathcal{G}$ is
strongly connected if there exists a path between any two distinct
vertices. A digraph $\mathcal{G}^f$ is a subgraph of $\mathcal{G}$
if its vertex set $\mathcal{V}(\mathcal{G}^f)\subseteq \mathcal{V}$,
arc set $\mathcal{E}(\mathcal{G}^f)\subseteq \mathcal{E}$ and every
arc in $\mathcal{E}(\mathcal{G}^f)$ has both end-vertices in
$\mathcal{V}$. A subgraph $\mathcal{G}^f$ is an induced subgraph if
two vertices of $\mathcal{G}^f$ are adjacent in $\mathcal{G}^f$ if
and only if they are adjacent in $\mathcal{G}$.  An induced subgraph
$\mathcal{G}^f$ that is strongly connected and maximal (i.e., no
more vertices can be added while preserving its connectedness) is
called a strong component of $\mathcal{G}$. In this paper we will
use the vertex set $\mathcal{V}(\mathcal{G}^f)=\{1,\cdots, n\}$ of
subgraph $\mathcal{G}^f$ to label the follower-agents. For a vertex
$i$ of $\mathcal{G}^f$, we call $\mathcal{N}_i=\{j:(i,j)\in
\mathcal{E}\}$ the neighbor set of vertex $i$. A nonnegative matrix
$A=[a_{ij}]\in \mathbf{R}^{n\times n}$ is called an  adjacency
matrix of subgraph $\mathcal{G}^f$ if the element $a_{ij}$
associated with the arc $(i,j)$ is positive, i.e. $a_{ij}=1
\Leftrightarrow (i,j) \in \mathcal{E}$. Moreover, we assume
$a_{ii}=0$ for all $i \in \mathcal{V}$. Notice that the adjacency
matrix $A$ may not be a symmetric matrix for a digraph. If
$\sum\limits_{j=1}^n a_{ij}=\sum\limits_{j=1}^n a_{ji}$ for
$i=1,\cdots, n,$ then the digraph $\mathcal{G}^f$ is called
balanced. A diagonal matrix $D=diag\{ d_1,...,d_n\}\in \mathbf{R}^{n
\times n}$ is called the degree matrix whose diagonal elements
$d_i=\sum\limits_{j=1}^n a_{ij}$ for $i=1,...,n$. Then the Laplacian
matrix of subgraph $\mathcal{G}^f$ is defined as
\begin{equation}
L=D-A,
\end{equation}
which may not be a symmetric matrix either. By this definition every
row sum of the Laplacian matrix $L$ is zero. Therefore, $L$ always
has a zero eigenvalue corresponding to a right eigenvector
$\textbf{1}=col(1,\cdots,1)\in \mathbf{R}^n$. Moreover, if subgraph
$\mathcal{G}^f$ is balanced, $L$ has a zero eigenvalue corresponding
to a left eigenvector $\textbf{1}\in \mathbf{R}^n$.

When the digraph $\mathcal{G}$ is used to describe the
interconnection topology of a multi-agent system consisting of one
active leader-agent and $n$ follower-agents, we can define a
diagonal matrix $B=diag\{a_{10},\cdots, a_{n0}\}\in \mathbf{R}^n$ to
be a leader adjacency matrix, where $a_{i0}=1$ if follower $i$ is
connected to the leader across the communication link $(i,0)$,
otherwise, $a_{i0}=0$.

If we define a new matrix $H=L+B\in \mathbf{R}^n,$ the following
lemma plays a key role in sequel.
\begin{lem} (\cite{hu07})\label{posit}
The following statements are equivalent:
\begin{enumerate}
\item[(1)] Vertex $0$ is a globally reachable vertex of digraph $\mathcal{G}$;
\item[(2)] $H$ is a positive stable matrix whose eigenvalues have positive real-parts;
\item[(3)] Furthermore, if $\mathcal{G}^f$ is balanced, $H+H^T$ is a symmetric positive definite matrix.
\end{enumerate}
\end{lem}

\begin{rem}
The topology connectedness in the sense that vertex $0$ is a
globally reachable vertex of digraph $\mathcal{G}$ implies that the
information of the leader can be propagated over the multi-agent
network. Obviously, this notion of connectedness is much weaker than
the notion of strong connectedness.
\end{rem}

For the purpose of modelling the time-variation of the
interconnection topology $\mathcal{G}$ of the leader-follower
multi-agent system, we adopt the following general assumptions:
\begin{itemize}
  \item[A1] There exists a switching signal $\sigma: [t_0,\infty)\rightarrow
  \mathcal{P}=\{1, 2, \cdots, N\}$, which is piecewise-constant. Here,
  $N$ denotes the total number of all possible interconnection topologies of the
  multi-agent system and $t_0$ is the initial time.
  \item[A2] If the time interval $[t_0, \infty)$ is constituted by
  an infinite sequence of bounded, non-overlapping, contiguous time-intervals
  $[t_j, t_{j+1})$ for $j=0,1,\cdots$ with $t_0=0$, there exists a
  positive constant $\tau$ such that $t_{j+1}-t_j\geq \tau$. The number $\tau$ is called
  a dwell time.
\end{itemize}
Then during each time-interval $[t_j,t_{j+1})$ the digraph
$\mathcal{G}_{\sigma(t)}$ is time-invariant and denoted by
$\mathcal{G}_p$ for some $p\in \mathcal{P}$.

\subsection{Leader-following problem}
In this paper we will study a distributed control design for a
leader-follower multi-agent
 system with one active leader-agent (just called leader in sequel for simplicity and labeled 0)
and $n$ cooperative follower-agents (just called followers in sequel
for simplicity). Consider a tracking control problem for a
multi-agent system where the followers are moving with the
first-order dynamics
\begin{equation}\label{follh}
\dot{x}_i(t)=u_i(t),
\end{equation}
for $i=1,\cdots, n$, and the dynamics of the leader is described by
the second-order differential equation
\begin{equation}
\label{leader}
\begin{cases} \dot{x}_0(t)=v_0(t),\\
\dot{v}_0(t)=a_0(t),\\
y_0(t)=x_0(t).
\end{cases}
\end{equation}

The variables $x_i(t), u_i(t)\in \mathbf{R}^{m}\;(i=1,\cdots, n)$
denote the states and inputs of $n$ followers respectively while
$x_0(t), v_0(t)\in \mathbf{R}^{m}$ and $a_0(t)\in \mathbf{R}^{m}$
denote the position, the velocity and the acceleration of the the
active leader respectively, and $y_0(t)$ is the only output. Here
for notation simplicity let $m=1$.

It was assumed in most existing works that an  information exchange
between agents is perfect, that is, each agent can obtain the
information of its neighbors precisely. In addition,  it was assumed
that the interconnection topology of the followers are undirected.
However, these assumptions are not valid in most practical
situations due to various reasons, such as sensor and/or
communication constraints, link variations. Measurement noises and
time-varying directed graph have to be considered for control of
leader-follower multi-agent systems.

Since the velocity $v_0(t)$ of the active leader cannot be measured
by followers, then each follower has to make estimation of $v_0(t)$
for control design by using the noisy measurements from its
neighbors. Our objective is to design a distributed control for the
leader-follower multi-agent system under partial and noisy
measurements and time-varying directed interconnection  topology so
that each follower can track the active leader and the velocity
estimation errors are convergent to zero in the sense of mean
square, i.e.,
\begin{equation}
\label{resl}
\begin{aligned}
\lim\limits_{t\to \infty}
E[(x_i(t)-x_0(t))^2]&=0,\\
 \lim\limits_{t\to \infty}
E[(v_i(t)-v_0(t))^2]&=0,
\end{aligned}
\end{equation}
where $v_i(t)$ is the estimate of $v_0(t)$ for the $i$th follower.
In this case, the closed loop system is said to be stochastically
stable in mean square.

\section{Distributed control of leader-follower system}\label{design}

In this section we will focus on designing a dynamic tracking
control for the leader-follower multi-agent system such that the
closed loop control system is stochastically stable in mean square.

The typical information available for each follower is its relative
position with its neighbors. However as mentioned in section of
introduction, the real information exchange among followers  through
a communication network is often subject to  different kinds of
constraints such as sensor noise, quantization errors, etc. In this
case, the  information available for the $i$th follower with respect
to its neighbors can be described as:
\begin{equation}
\label{meas}
z_{ij}(t)=a_{ij}(t)(x_i(t)-x_j(t)+\varrho_{ij}\omega_{ij}(t))\in
\mathbf{R},
\end{equation}
where  $j\in\mathcal{N}_i(t)$  with $\mathcal{N}_i(t)$ being the
neighbor set of follower $i$ at time $t$, $a_{ij}(t)$ is the
connection weight between agent $i$ and agent $j$ at time $t$,
$\omega_{ij}(t)$ is an independent normal white noise,
$\varrho_{ij}\geq 0$ is the noise intensity.

It is noted that since only the relative noisy position measurements
$z_{ij}(t)$  can be used for the $i$th follower, the construction of
a distributed estimator  and controller turns out to be much  more
challenging than that in \cite{hong06}. To address the challenge, a
novel decomposition scheme of the velocity $v_0(t)$ of the active
leader is proposed as follows:
\begin{equation}
\label{deco}
\begin{cases}v_0(t)=\alpha(t) \mathbf{v}_0(t),\\
\dot{\mathbf{v}}_0(t)=\mathbf{a}_0(t),
\end{cases}
\end{equation}
where $\mathbf{v}_0(t)$ is a  continuous differentiable function
called nominal velocity and $\alpha(t):[t_0,\infty)\to (0,\infty)$
is a continuous differentiable function satisfying
$\int_{t_0}^{\infty}\alpha(s)ds=\infty$ and
$\int_{t_0}^{\infty}\alpha^2(s)ds<\infty$. In addition, $\alpha(t)$
has an upper bound $\mu$ in $[t_0, \infty)$. We call
$\mathbf{a}_0(t)$ the nominal acceleration. Then the relationship
between the acceleration $a_0(t)$ and the nominal one
$\mathbf{a}_0(t)$ can be expressed as
\begin{equation}\label{acce}
a_0(t)=\dot{\alpha}(t)\mathbf{v}_0(t)+\alpha(t)\mathbf{a}_0(t).
\end{equation}
Notice that $\alpha(t)$ in the decomposition (\ref{deco}) can be
easily found for a continuous differentiable function $v_0(t)$, for
example, $\alpha(t)=\frac{1}{t+1}$ with its upper bound $\mu=1$ in
time-interval $[0,\infty)$. In sequel we assume $\alpha(t)$ and
$\mathbf{a}_0(t)$ are precisely known beforehand.

\begin{rem}
Let $v_i(t)=\alpha(t)\mathbf{v}_i(t)$ be the estimate of $v_0(t)$
by the $i$th follower. If $\mathbf{v}_i(t)-\mathbf{v}_0(t)\to 0$,
one has $v_i(t)-v_0(t)\to 0$ since $\alpha(t)$ has an upper bound
$\mu$ during time-interval $[t_0, \infty)$.
\end{rem}

On the basis of the decomposition (\ref{deco}), for the $i$th
follower with dynamics (\ref{follh}), we propose the following local
dynamic control scheme with an estimator:
\begin{equation}
\label{cont}\begin{cases} u_i(t)=-k\alpha(t) \sum\limits_{j\in
\mathcal{N}_i(t)}z_{ij}(t)+\alpha(t) \mathbf{v}_i(t),\\
\dot{\mathbf{v}}_i(t)=\mathbf{a}_0(t)-\gamma k\alpha(t)
\sum\limits_{j\in \mathcal{N}_i(t)}z_{ij}(t),
\end{cases}
\end{equation}
where $\mathbf{v}_i(t)$ is an estimate of the nominal velocity
$\mathbf{v}_0(t)$ for $i=1,\cdots,n,$ $0<\gamma<1,$  the gain
constant $k>0$ is to be determined in sequel.

\begin{rem}
It is noted that the  estimator for the nominal velocity
$\mathbf{v}_0(t)$ of the active leader is a distributed one based
 on measurements of relative positions of its neighbors. The rationale for the estimator
 is to collect the position information of the leader within the neighborhood
  of the $i$th follower during a time period and then make a tendency prediction
   of the trajectory of the leader with the gathered historical data through an integrator.
\end{rem}

In the dynamic control (\ref{cont}) the neighbor set
$\mathcal{N}_i(t)$ at time $t$ of the $i$th follower may include the
active leader. We divide the neighbor set $\mathcal{N}_i(t)$ into
two subsets as follows:
\begin{equation}
\label{nebr} \mathcal{N}_i(t)=\mathcal{N}_i^f(t) \cup
\mathcal{N}_i^l(t), i=1,\cdots, n,
\end{equation}
where $\mathcal{N}_i^f(t)$ denotes the follower-neighbor set and
$\mathcal{N}_i^l(t)$ denotes the leader-neighbor set of follower
$i$.
 Then applying the dynamic control scheme (\ref{cont}) to  system (\ref{follh}) yields:
\begin{equation}
\label{flexp}
\begin{aligned}
\dot x_i(t)=&-k\alpha(t)\{\sum\limits_{j\in
\mathcal{N}_i^f(t)}a_{ij}(t)(x_i(t)-x_j(t))\\
&+a_{i0}(t)(x_i(t)-x_0(t))\}-k\alpha(t)a_{i0}(t)\varrho_{i0}\omega_{i0}(t)\\
&-k\alpha(t)\sum\limits_{j\in
\mathcal{N}_i^f(t)}a_{ij}(t)\varrho_{ij}\omega_{ij}(t)+\alpha(t) \mathbf{v}_i(t),\\
\dot{\mathbf{v}}_i(t)=&\mathbf{a}_0(t)-\gamma
k\alpha(t)\{\sum\limits_{j\in
\mathcal{N}_i^f(t)}a_{ij}(t)(x_i(t)-x_j(t))\\
&+a_{i0}(t)(x_i(t)-x_0(t))\}-\gamma k\alpha(t)a_{i0}(t)\varrho_{i0}\omega_{i0}(t)\\
 &-\gamma k\alpha(t)\sum\limits_{j\in
\mathcal{N}_i^f(t)}a_{ij}(t)\varrho_{ij}\omega_{ij}(t).
\end{aligned}
\end{equation}
Let $a(i,\cdot)$ denote the $i$th row of the  adjacency matrix
$A=[a_{ij}]\in \mathbf{R}^{n\times n}$ of digraph $\mathcal{G}^f$.
Denote $x=col(x_1,\cdots,x_n)\in\mathbf{R}^{n},
\mathbf{v}=col(\mathbf{v}_1,\cdots,\mathbf{v}_n)\in\mathbf{R}^{n},$
$\omega_0=col(\omega_{10},\cdots,$ $ \omega_{n0})\in\mathbf{R}^{n},$
$\omega_i=col(\omega_{i1},\cdots, \omega_{in})\in\mathbf{R}^{n}$ for
$i=1,\cdots, n$ and $\omega=col(\omega_0,\omega_1, \cdots,
\omega_n)\in\mathbf{R}^{n(n+1)}$. Then system (\ref{flexp}) can be
rewritten in a compact form:
\begin{equation}
\label{flc}
\begin{cases}
\dot x=-k\alpha H_{\sigma}x+k\alpha B_{\sigma}\mathbf{1}x_0-k\alpha\Sigma_\sigma \omega+\alpha \mathbf{v},\\
\dot{\mathbf{v}}=\mathbf{a}_0\mathbf{1}-\gamma k \alpha
H_{\sigma}x+\gamma k\alpha B_{\sigma}\mathbf{1}x_0-\gamma
k\alpha\Sigma_\sigma \omega,
\end{cases}
\end{equation}
where $\sigma$ is the piecewise-constant switching signal,
$H_{\sigma}=L_{\sigma}+B_{\sigma}$, $L_{\sigma}$ is the Laplacian
matrix associated with the switched subgraph
$\mathcal{G}_{\sigma}^f$,$B_{\sigma}$ is the leader adjacency matrix
associated with the switched digraph $\mathcal{G}_\sigma$,
$\Sigma_0=diag\{\varrho_{10},\cdots, \varrho_{n0}\}, $
$\Sigma_i=diag\{\varrho_{i1},\cdots, \varrho_{in}\}$ for $i=1,
\cdots, n$, and the matrix $\Sigma_{\sigma}$ is defined in equation
(\ref{sigma}).

\begin{table*}[hbt]
\begin{equation}\label{sigma}
\begin{aligned}
\Sigma_{\sigma}=&\begin{pmatrix}a_{10}\varrho_{10}&&&a_{11}\varrho_{11}&\cdots &a_{1n}\varrho_{1n}& & & &\\
&\ddots& & &\ddots&\ddots&\ddots& & &\\
& & a_{n0}\varrho_{n0}& & & & &a_{n1}\varrho_{n1} &\cdots&a_{nn}\varrho_{nn}\end{pmatrix}\in \mathbf{R}^{n\times(n(n+1))}\\
=&[B_{\sigma}\Sigma_0\;\;
diag\{a(1,\cdot)\Sigma_1,\cdots,a(n,\cdot)\Sigma_n\}]\in R^{n\times
n(n+1)}.
\end{aligned}\end{equation}
\end{table*}

In order to show that all the followers can track the active leader,
we firstly make two variable changes $\bar x=x-x_0\textbf{1}$ and
$\bar{\mathbf{v}}=\mathbf{v}-\mathbf{v}_0 \textbf{1}$. According to
the spectrum properties of graph Laplacian matrix, $L_\sigma
\mathbf{1}=0$ and then
$$\begin{aligned}-H_{\sigma}x+ B_{\sigma}\mathbf{1}x_0
=-H_{\sigma}\bar{x}.\end{aligned}$$ With system (\ref{leader}) and
(\ref{flc}), we have
\begin{equation}
\label{errs}
\begin{cases}
\dot{\bar x}=-k\alpha H_{\sigma}\bar x-k\alpha\Sigma_{\sigma} \omega+\alpha\bar{\mathbf{v}},\\
\dot{\bar{\mathbf{v}}}=-\gamma k \alpha H_{\sigma}\bar{x}-\gamma k\alpha\Sigma_{\sigma} \omega,\\
\end{cases}
\end{equation}
which can be rewritten in a compact form:
\begin{equation}
\label{errc} \dot{\varepsilon}=F_\sigma \varepsilon +\Omega_\sigma
\omega,
\end{equation}
where $\varepsilon=\begin{pmatrix}\bar
x\\\bar{\mathbf{v}}\end{pmatrix}, F_{\sigma}=\begin{pmatrix}-k
\alpha H_\sigma & \alpha I\\-\gamma k \alpha H_\sigma &
0\end{pmatrix},$ and
$\Omega_\sigma=\begin{pmatrix}-k \alpha \Sigma_{\sigma} \\
-\gamma k \alpha \Sigma_{\sigma}\end{pmatrix}.$

In sequel we will analyze the stochastic stability of system
(\ref{errc}). Two cases: time-invariant leader-follower topology and
time-varying leader-follower topology will be considered.

\subsection{Time-invariant topology}\label{fix}

When the leader-follower interconnection topology
$\mathcal{G}_{\sigma(t)}$ is time-invariant, the subscript
$\sigma(t)$ will be dropped.

Here we give a main result as follows.
\begin{thm}\label{thm0} If vertex $0$ is globally reachable in $\mathcal{G}$,
then with the dynamic tracking control (\ref{cont}) each follower
can track the active leader  asymptotically in mean square, that is,
\begin{equation*}
\begin{aligned}
&\lim\limits_{t\to \infty}
E[(x_i(t)-x_0(t))^2]=0,\\
&\lim\limits_{t\to \infty} E[(v_i(t)-v_0(t))^2]=0.
\end{aligned}
\end{equation*}
\end{thm}

Proof: To facilitate analysis, we write system (\ref{errc}) in the
form of It$\rm\hat{o}$ stochastic differential equation:
\begin{equation}
\label{errci0} d \varepsilon=F \varepsilon dt +\Omega d \rm w,
\end{equation}
where $\rm w(t)$ is an $n(n+1)$-dimensional standard Brownian
motion.

Choose a nonnegative function
\begin{equation}\label{lyap}
V(t)=\varepsilon^T(t) P \varepsilon(t),
\end{equation}
where
\begin{equation}\label{matrixp0}
P=\begin{pmatrix}\bar{P}&-\gamma \bar{P}\\
-\gamma \bar{P}&\bar{P}\end{pmatrix}
\end{equation}
and $\bar{P}$ is a symmetric positive definite matrix satisfying
$H^T\bar{P}+\bar{P}H=I_n$ which is well defined due to Lyapunov
Theorem and Lemma \ref{posit}.

It follows from the definition of $P$ in (\ref{matrixp0}) and $F$ in
(\ref{errc}) that
\begin{equation}
\label{lve0}
\begin{aligned}
&PF+F^TP=:-Q\\
&=-\alpha(t)\begin{pmatrix}
k(1-\gamma^{2})I_n&-\bar{P}\\
-\bar{P}&2\gamma \bar{P}
\end{pmatrix}.
\end{aligned}
\end{equation}
If we choose
\begin{equation}\label{gain0}
 k> \frac{\lambda_{\rm max}(\bar{P})}{2\gamma(1-\gamma^{2})},
\end{equation}
according to Schur complement formula, it can be shown that $Q$ is
positive definite.

By It$\rm\hat{o}$ formula, we  have
\begin{equation}
\label{lv0}
\begin{aligned}
dV(t)|_{(\ref{errci0})}=&[\varepsilon^T(t)(PF+F^TP)\varepsilon(t)+tr(P\Omega \Omega^T)]dt\\
&+2\varepsilon^T(t)P\Omega d\rm w(t)\\
=&[-\varepsilon^T(t)Q\varepsilon(t)+tr(P\Omega \Omega^T)]dt\\
&+2\varepsilon^T(t)P\Omega d\rm w(t)\\
\leq & \frac{-\lambda_{\rm min}(Q)}{(1+\gamma)\lambda_{\rm max}(\bar{P})}\alpha(t)V(t)dt+\rho_0 \alpha^2(t)dt\\
&+2\varepsilon^T(t)P\Omega d\rm w(t),\\
\end{aligned}
\end{equation}
where $\rho_0= n\lambda_{\rm
max}(\bar{P})k^2(1-\gamma^2)\max\limits_{\sigma\in
\mathcal{P}}\|\Sigma_\sigma\|^2$.

For the third term in the last inequality of (\ref{lv0}), we will
prove that the mathematical expectation
\begin{equation}
\label{intine0} E[\int_{t_0}^{t} \varepsilon^T(s)P\Omega d\rm
w(s)]=0,
\end{equation}
 for all $t \geq t_0$.

For any $t_0\geq 0$, $T\geq t_0$,  let
$\tau_{\delta}^{t_0}=\inf\{t\geq t_0: V(t)\geq \delta\}$ where
$\delta$ is a given positive number if $V(t)\geq \delta$ for some
$t\in [t_0,T]$; otherwise, $\tau_{\delta}^{t_0}=T$. From equation
(\ref{lv0}), one can get
\begin{equation}
\label{elv0}
\begin{aligned}
&E[V(t\wedge \tau_{\delta}^{t_0})\chi_{t\leq
\tau_\delta^{t_0}}]-E[V(t_0)]\\
\leq & \frac{-\lambda_{\rm min}(Q)}{(1+\gamma)\lambda_{\rm
max}(\bar{P})}\int_{t_0}^{t}\alpha(s)V(s\wedge
\tau_{\delta}^{t_0}))\chi_{s\leq \tau_\delta^{t_0}}ds\\
& +\rho_0\int_{t_0}^{t}\alpha^2(s)ds\\
\leq & \rho_0\int_{t_0}^{T}\alpha^2(s)ds,
\end{aligned}
\end{equation}
which implies that there exists a constant $\Delta_{t_0,T}$ such
that $E[V(t\wedge \tau_{\delta}^{t_0})\chi_{t\leq
\tau_\delta^{t_0}}]\leq \Delta_{t_0,T}, \;\forall t\in [t_0, T]$.
Then, by Fatou lemma \cite{chow}, we have
$$
\sup\limits_{t_0\leq t \leq T}E[V(t)]\leq  \Delta_{t_0,T}.
$$
Thus, $$E[\int_{t_0}^{t}\alpha^2(s) V(s)ds] \leq
\sup\limits_{t_0\leq t \leq T}E[V(t)] \int_{0}^{T}\alpha^2(s) ds<
\infty.$$ In addition, we have
$$
E[\int_{t_0}^{t}\|\varepsilon^T(s)P\Omega_{\sigma} \|^2ds] \leq
\rho_0 E[\int_{t_0}^{t}\alpha^2(s)V(s)ds].
$$

By the It$\rm\hat{o}$ integral formula \cite{frie}, we have the
equation (\ref{intine0}).

Then based on equations  (\ref{lv0}) and (\ref{intine0}), we have
\begin{equation*}
\label{elvt}
\begin{aligned}
&E[V(t)]-E[V(t_0)]\\
\leq & \frac{-\lambda_{\rm min}(Q)}{(1+\gamma)\lambda_{\rm
max}(\bar{P})}\int_{t_0}^{t}\alpha(s)E[V(s)]ds
+\rho_0\int_{t_0}^{t}\alpha^2(s)ds.
\end{aligned}
\end{equation*}

Then, by the comparison theorem \cite{mich}, we have
$$
\begin{aligned}
&E[V(t)]\leq E[V(t_0)]\exp\{\frac{-\lambda_{\rm
min}(Q)}{(1+\gamma)\lambda_{\rm
max}(\bar{P})}\int_{t_0}^t\alpha(s)ds
\}\\
&+\rho_0 \int_{t_0}^{t}\alpha^2(s) \exp\{\frac{-\lambda_{\rm
min}(Q)}{(1+\gamma)\lambda_{\rm max}(\bar{P})}\int_{s}^t\alpha(\iota)d\iota\}ds\\
&\leq E[V(t_0)]\exp\{\frac{-\lambda_{\rm
min}(Q)}{(1+\gamma)\lambda_{\rm
max}(\bar{P})}\int_{t_0}^t\alpha(s)ds
\}\\
&+\rho_0 \int_{t_0}^{t}\alpha^2(s)ds.
\end{aligned}
$$
Therefore, when $t\to \infty,$ $\int_{t_0}^{t}\alpha(s)ds\to \infty,
\int_{t_0}^{t}\alpha^2(s)ds\to 0$, and so $E[V(t)]\to 0$.

Since $V(t)\geq (1-\gamma)\lambda_{\rm
min}(\bar{P})\|\varepsilon(t)\|^2$, one has
$$
\begin{aligned}
\lim\limits_{t\to \infty}E(x_i(t)-x_0(t))^2=0,\\
\lim\limits_{t\to \infty}E(\mathbf{v}_i(t)-\mathbf{v}_0(t))^2=0.
\end{aligned}
$$
It then follows from $0\leq \alpha(t)\leq \mu$ that
$\lim\limits_{t\to \infty}E(\alpha(t)\mathbf{v}_i(t)-v_0(t))^2=0$.
The proof is thus completed. \hfill\rule{4pt}{8pt}

\begin{rem}
From the proof of Theorem \ref{thm0} it can be seen that the
introduction of the gain function $\alpha(t)$ can ensure equation
(\ref{intine0}) and $E[V(t)]\to 0$ as $t \to \infty$. Thus the
tracking result presented in this paper is much more improved  in
comparison with  that in \cite{hong06} for the leader-follower
multi-agent system with directed interconnection topology.
\end{rem}

\subsection{Time-varying topology}\label{swit}

Now one is ready to present the following main result about
leader-follower tracking control under time-varying interconnection
topology.

\begin{thm} \label{thm} If vertex $0$ is globally reachable in
$\mathcal{G}_\sigma$ and $\mathcal{G}_\sigma^{f}$ is balanced during
each time-interval $[t_j, t_{j+1})$, then with the dynamic tracking
control (\ref{cont}) each follower can track the active leader
asymptotically in mean square.
\end{thm}

Proof: Take a nonnegative function $V(t)=\varepsilon^T(t) P
\varepsilon(t)$ with symmetric positive definite matrix
\begin{equation}
\label{matrixp} P= \left(\begin{array}{cc}I_n&-\gamma I_n\\-\gamma
I_n&I_n\end{array}\right).
\end{equation}

It follows from the definition of $P$ in (\ref{matrixp}) and
$F_\sigma$ in (\ref{errc}) that
\begin{equation}
\label{lve}
\begin{aligned}
&PF_\sigma+F_\sigma^TP=:-Q_\sigma\\
&=-\alpha(t)\begin{pmatrix}
k(1-\gamma^{2})  (H_{\sigma}+H_{\sigma}^T)&-I_n\\
-I_n&2\gamma I_n
\end{pmatrix}.
\end{aligned}
\end{equation}
By assumptions in Theorem \ref{thm} and Lemma \ref{posit}, $H_\sigma
+H_{\sigma}^T$ is positive definite. If we choose
\begin{equation}\label{gain}
 k> \frac{1}{2\gamma(1-\gamma^{2})\bar{\lambda}},
\end{equation}
where $\bar{\lambda}=\min\limits_{\sigma\in
\mathcal{P}}\{\lambda_\sigma: \;\mbox{eigenvalues of}\;
H_\sigma+H_{\sigma}^T \}>0$, according to Schur complement formula,
it can be shown that $Q_\sigma$ is positive definite.

The rest of the proof are similar to those in Theorem \ref{thm0} by
noting that $\mathcal{P}$ is a finite set, and hence omitted.
\hfill\rule{4pt}{8pt}

\begin{rem}
In Theorem \ref{thm}, the condition that $\mathcal{G}_\sigma^f$ is
balanced is a sufficient condition. The subsequent numerical example
shows that this condition is not necessary for the mean square
convergence of  the tracking errors.
\end{rem}

\section{A simulation example}\label{simu}
\begin{figure}[!htp]
\centering \subfigure[$\mathcal{G}_1$ and $\mathcal{G}_1^f$]
{\includegraphics[width=0.15\textwidth]{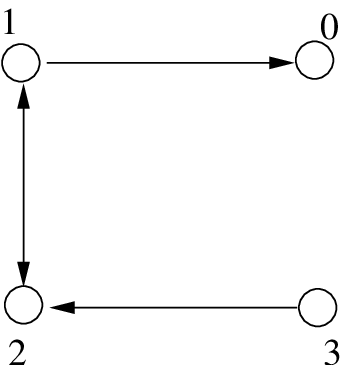}}
\mbox{\hspace{1cm}} \subfigure[$\mathcal{G}_2$ and
$\mathcal{G}_2^f$]
{\includegraphics[width=0.15\textwidth]{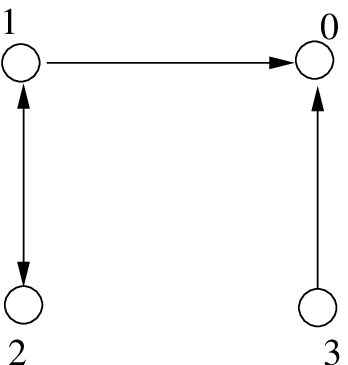}}
\caption{$\mathcal{V}=\{0,1,2,3\}$ and
$\mathcal{V}(\mathcal{G}^f)=\{1,2,3\}$}\label{fig1}
\end{figure}

In this section a numerical example is given to illustrate the
proposed dynamic tracking control algorithm. Consider a
leader-follower multi-agent system with one active leader and three
followers. Suppose that the leader-follower interconnection topology
$\mathcal{G}_\sigma$ is time-varying with switching rule:
$\mathcal{G}_1,\mathcal{G}_2,\mathcal{G}_1,\mathcal{G}_2,\cdots,$
where $\mathcal{G}_1$ and $\mathcal{G}_2$ are described in Fig.
\ref{fig1}.
Then one has the following Laplacian matrices
$$L_1=\begin{pmatrix}1&-1&0\\
-1&1&0\\0&-1&1\end{pmatrix},
L_2=\begin{pmatrix}1&-1&0\\
-1&1&0\\0&0&0\end{pmatrix},$$ and the leader adjacency matrices
$B_1=diag\{1,0,0\}$ and $B_2=diag\{1,0,1\}$. It is not difficult to
have the minimal eigenvalue $\bar{\lambda}=0.3187$ for $H_1=L_1+B_1$
and $H_2=L_2+B_2$.

\begin{figure}[!htp]
\centering
\includegraphics[width=\columnwidth]{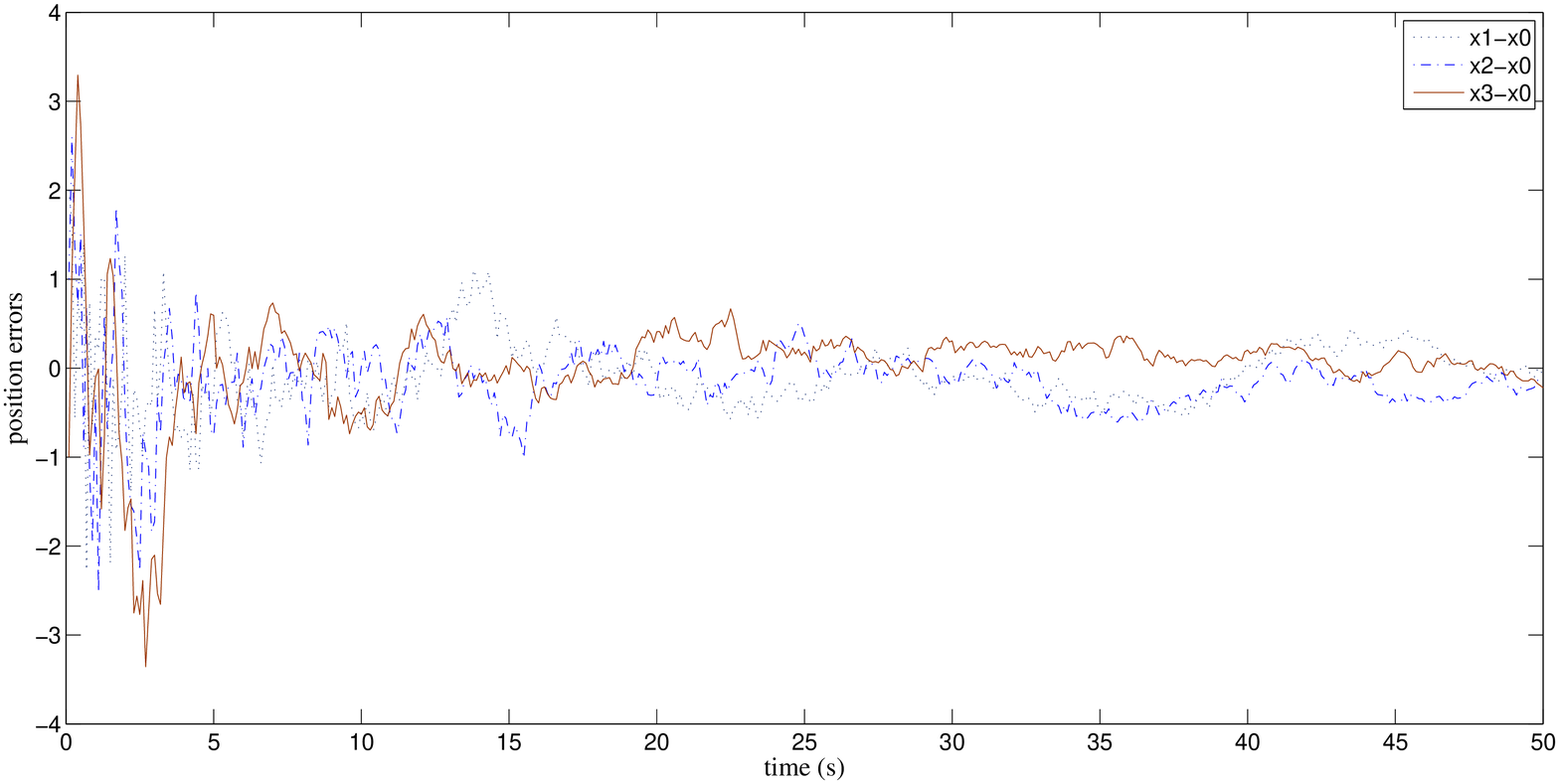}
\caption{The evolution of tracking errors}
 \label{tracfig}
\end{figure}

In the control (\ref{cont}) we choose $\alpha(t)=\frac{1}{t+1}$,
$\gamma=0.8$ and $k=6$. In addition, let the intensity
$\varrho_{ij}=1$ when $a_{ij}=1$. For system (\ref{errc}), the
initial value of $\varepsilon(t)$ is taken randomly as
$col(2,1,-1,-0.2,-2,0.2)$.  Then the tracking errors $\bar{x}_1(t),
\bar{x}_2(t)$ and $\bar{x}_3(t)$ are shown in Fig. \ref{tracfig}. It
can be seen that the tracking control (\ref{cont}) ensures that the
followers track the active leader under noisy measurements. Notice
that even though digraph $\mathcal{G}_1^f$ is not balanced, the
tracking errors still converge in mean square.

\section{Conclusions}\label{con}
In this paper we have studied the leader following problem of a
multi-agent system with measurement noises and directed
interconnection topology. The neighbor-based distributed control
scheme with distributed estimators has been developed. Algebraic
graph theory and stochastic analysis have been employed to analyze
the mean square convergence of the tracking errors. One possible
future research topic is to study the leader-following problem in a
noisy environment when the dynamics of each agent is described by a
more general linear system.

\section*{Acknowledgements}

The authors are most grateful to the associate editor and reviewers
for their many constructive comments based on which  this paper has
been significantly improved. This work was partially supported by a
grant from City University of Hong Kong under Grant No. 9360131.

\balance
\bibliography{autosam}

\begin{thebibliography}{99}
\bibitem[Wang$\;$P. K. C., 1991]{wang} Wang P. K. C. (1991). Navigation strategies for multiple autonomous mobile
robots moving in formation, {\it J. Robot. Syst.}, vol. 8, no. 2,
177-195.

\bibitem[Das$,$Fierro$,\&$Kumar, 2002]{das} Das A. K., Fierro R., $\&$ Kumar V., et al.
(2002). A vision-based  formation  control  framework, {\it IEEE
trans. Robot. Autom.}, vol. 18, no. 5, 813-825.

\bibitem[Vanek$,$Peni$,$et al., 2005]{vane} Vanek B.,  Peni T., Bokor J., $\&$ Balas G. (2005). Practical approach to
real-time trajectory tracking of UAV formations, in {\it Proc. of
American Control Conference}, Oregon,  122-127.

\bibitem[Anderson$,$Fidan$,$et al., 2008]{anders} Anderson B. D. O., Fidan B., Yu C., $\&$ Walle D. (2008). UAV formation
control: theory and application, {\it Lecture Notes in Control and
Information Sciences}, Springer: Berlin, 371, 15-33.

\bibitem[Gupta$,$Cao$,\&$Haering, 2008]{gupta} Gupta H., Cao X., $\&$ Haering N. (2008). Map-based active
leader-follower surveillance system, in Proc. of {\it ECCV workshop
on Multi-Camera and Multi-modal Sensor Fusion Algorithms and
Applications}, Marseille, France.

\bibitem[Hu$,\&$Hu, 2008]{hu} Hu J., $\&$ Hu X. (2008). Optimal target trajectory estimation and filtering
using networked sensors. {\it Jr. Systems Science $\&$ Complexity},
vol. 21, 325-336.

\bibitem[Jadbabaie$,$Lin$,\&$Morse, 2003]{jad} Jadbabaie A.,  Lin J., $\&$ Morse A. S. (2003). Coordination of groups of
mobile autonomous agents using nearest neighbor rules, {\it IEEE
Trans. on Automatic Control}, vol. 48, no. 6, 988-1001.

\bibitem[Ren$,\&$Beard, 2005]{ren} Ren W. $\&$ Beard R. W. (2005). Consensus seeking in multiagent systems
under dynamically changing interaction topologies, {\it IEEE Trans.
on Automatic Control},  vol. 50, no. 5, 655-661.

\bibitem[Lin$,$Francis$,\&$Maggiore, 2005]{lin} Lin Z., Francis B., $\&$ Maggiore M. (2005). Necessary and sufficient graphical conditions for
formation control of unicycles, {\it IEEE Trans. on Automatic
Control},  vol. 50, no. 1, 121-127.

\bibitem[Shi$,\&$Hong, 2009]{shi} Shi G., $\&$ Hong Y. (2009). Global target aggregation and state
agreement of nonlinear multi-agent systems with switching
topologies, {\it Automatica}, vol. 45, no. 5, 1165-1175.


\bibitem[Hu$,\&$Hong, 2007]{hu07} Hu J., $\&$ Hong Y. (2007). Leader-following coordination of multi-agent systems with coupling
time delays, {\it Physica A},  vol. 374, no. 2, 853-863.

\bibitem[Lin$,$Jia$,$ et al., 2008]{lin08} Lin P., Jia Y., Du J., $\&$ Yuan S. (2008). Distributed control of multi-agent systems with second-order agent
dynamics and delay-dependent communications, {\it Asian J. Control},
vol. 10, no. 2, 254-259.

\bibitem[Fax$,\&$Murray, 2004]{fax} Fax A., $\&$ Murray R. M., (2004). Information flow and cooperative
control of vehicle formations. {\it IEEE Trans. on Automatic
Control}, vol. 49, no. 9, 1465-1476.

\bibitem[Hong$,$Hu$,\&$Gao, 2006]{hong06} Hong Y., Hu J.,$\&$, Gao L. (2006). Tracking control for multi-agent consensus with an
active leader and variable topology, {\it Automatica}, vol. 42, no.
7, 1177-1182.

\bibitem[Li$,\&$Zhang, 2009]{li09} Li T., $\&$ Zhang J. F. (2009).  Mean square average consensus under
measurement noises and fxed topologies: necessary and sufficient
conditions, {\it Automatica}, vol. 45, no. 8, 1929-1936.

\bibitem[Huang$,\&$Manton, 2009]{huang} Huang M., $\&$ Manton J. H. (2009), Coordination and consensus of networked
agents with noisy measurement: stochastic algorithms and asymptotic
behavior, {\it SIAM Journal on Control and Optimization}, vol. 48,
no. 1, 134-161.

\bibitem[Godsil$,\&$Royle, 2001]{cgod}  Godsil C., $\&$ Royle G. (2001). {\it Algebraic
Graph Theory,} New York: Springer-Verlag.

\bibitem[Nevelson$,\&$Hasminskii, 1976]{neve}  Nevelson M. B.,$\&$ Hasminskii R. Z. (1976). {\it Stochastic
Approximation and Recursive Estimation},  Providence : American
Mathematical Society.


\bibitem[Chow$,\&$Teicher, 1997]{chow}  Chow Y. S., $\&$ Teicher H. (1997). {\it Probability theory: independence, interchangeability,
martingales}, New York: Springer.

\bibitem[Friedman$\;$A., 1975]{frie} Friedman A. (1975). {\it Stochastic differential equations and
applications: Vol. 1}, New York: Academic Press.

\bibitem[Michel$,\&$Miller, 1977]{mich}  Michel A. N., $\&$ Miller R. K. (1977) {\it Qualitative analysis of large
scale dynamical systems}, New York: Academic Press.

\end{thebibliography}
\bibliographystyle{harvard}

\end{document}